\documentclass[showpacs,preprintnumbers,amsmath,amssymb] {revtex4}
\usepackage{graphicx,color,epsfig}
\usepackage{dcolumn}
\usepackage{bm}

\begin{document}
\topmargin 0.2cm

\title{ \bf Effect of two
loop correction in the formation of QGP droplet}

\author{S. Somorendro Singh\footnote{email:sssingh@physics.du.ac.in} }
\affiliation {Department of Physics and Astrophysics, University of Delhi, Delhi - 110007, India \\
}        

\begin{abstract}
 The effect of two loop correction in the formation
of quark-gluon plasma (QGP) 
droplet is studied with the introduction of
the two loop correction factor 
in the mean field potential. Due to the correction factor
it shows stability in the droplet formation of QGP indicating
at different parametrization factors of the QGP fluid. 
The correction factor in the potential 
also shows gluon parameter factor
shifts to a larger value from its earlier value of 
gluon factor of one loop correction in obtaining the
stable droplets. The 
results show decreasing in the observable QGP droplets and droplet
sizes are found to be $1.5-2.0$ fm radii with the two loop correction. 
It indicates that there is parameter like 
Reynold's number which can control the dynamics of QGP 
droplet formation and the stability of droplet in the case of droplet
formation with the two loop correction factor. 
\end{abstract}
\maketitle

\section{Introduction}

 Lattice theory indicates about the 
phase transition~\cite{ali}
from a deconfined phase of free quarks and gluons to a confined
phase of hadrons. The deconfined state of matter is broadly known as
quark-gluon plasma (QGP), that is probably obtained
in the early universe formation. The study has made the theorist and 
experimentalist highly busy in the last two decades in search of
identifying the formation of QGP. During these decades, 
numbers of highly energetic laboratories are set up around the globe and 
focusing to find out how the early universe started to create. They believe 
that matter present in the earlier time was as deconfined of
free quarks and gluons known as quark-gluon plasma (QGP), and if it is formed
then it would expand hydrodynamically
with subsequent cooling lead to
formation of confined colorless matter of quarks of hadrons. 
So the process of 
early universe creation is
considered to be a complicated phenomena and this complicated 
nature indicates the study of QGP fireball in Ultra 
Relativistic Heavy-Ion Collisions an exciting field in the present day of
heavy ion collider physics~\cite{satz}. There are a number 
of phenomenological methods which
try to solve these complicated phenomena. We also try
to solve the problem by considering a phenomenological potential model. 
In the model we try to create free energy evolution through the different
quark and gluon flow parameters forming various
sizes of droplet. The formation of droplet differs
with the change of temperature and somehow with the parametrization
values can make a few stable droplets. This indicates that 
droplet formation with the paramatrization value of quark and gluon
is dependent on the temperature. The droplets
determine the critical size when transforming the phase from
a quark-gluon to a confined phase of hadron droplets.
On the basis of the critical radius
of the droplet, we also calculate the surface tension
considered to be a parameter making the sharp
boundary between these two phases. Moreover 
the calculation of surface tension gives another important
property of liquid drop model in determining the stability 
of droplet formation in the system.
\par So, in this paper, we focus 
on the stability of QGP droplet formation through the different 
quark and gluon parameters
incorporating the two loop correction in mean field potential.
To evaluate the
droplet formation, we use the thermodynamic partition function 
with the correction of 
two loop potential in the Hamiltonian of the system.
The partition function is correlated through Gibbs free energy which
is developed through density of state~\cite{mardor,neergaard}.
The density of state can be established through the earlier process of Thomas 
and Bethe
model incorporating the one and two loop correction in the potential.
The correction in the potential with loops affect in the droplet
size and impacts in 
the stability of
droplet formation with the variation of the dynamical quark and gluon flow 
parameters.
\par The paper is organised as: In section $II$, it briefly
tries to construct the Hamiltonian of the system
incorporating one loop correction extending to the two 
loop correction factor in the potential and set up the
Gibbs free energy of the two loop correction. In section $III$ 
 the free energy evolution of system is discussed. In section $IV$ it 
presents the surface tension of the system symbolising the stable
droplet formation of QGP. In section $V$, the analytical
solutions as results are discussed. At last, the conclusion with the
details of stable droplet formation of QGP with different 
flow parametrization
values is presented.

\section{Hamiltonian and mean field potential with two loop correction}
 The dynamical behaviour of quarks-antiquarks and gluons in QGP enforce
us in identifying the interacting potential among quark-quark,quark-antiquark, 
quark and gluon, which in turn, give the bulk thermodynamical 
and hydrodynamical
properties of the particles of the system. So the effective mean field potential for QGP
is calculated through the thermal mass formalism and the corresponding thermal
Hamiltonian leading to the confining/de-confining potential among them.
The thermal -Hamiltonian is obtained as~\cite{peshier,rama}:
\begin{eqnarray}
H(k,T) &=&[k^{2}+m^{2}(T)]^{1/2} \nonumber\\
       &=& k+m^{2}(T)/2k~~for~large~k  
\end{eqnarray}
\begin{equation}
H(k,T)=k+m_{0}^{2}/2k-\{m_{0}^{2}-m^{2}(T)\}/2k
\end{equation}
where ,
\begin{equation} m^{2}(T)=\frac{16 \pi}{k}\gamma_{q,g} ~ \alpha_{s}(k) T^{2} [1+\frac{\alpha_{s}(k)}{4\pi} a_{1}+\frac{\alpha_{s}^{2}(k)}{16 \pi^2}a_{2} ].
\end{equation} It is thermal mass obtained after
one and two loop corrections being introduced in the potential.
The co-efficients used in the thermal mass $a_{1}$ and $a_{2}$ are the one 
and two
loop correction factors which are obtained through the interactions among
the constituent
particles. They are defined numerically depending on the number
of quark flavours and they are given as:
\begin{eqnarray}
a_{1}&=&2.5833-0.2778~ n_{l}, \\
a_{2}&=&28.5468-4.1471~n_{l}+0.0772~n_{l}^{2}
\end{eqnarray}
where $n_{l}$ is considered to be the number of light quark 
elements~\cite{fischler,smirnov}.
$k$ is the quark (gluon) 
momentum and $m_{0}$, the dynamic rest mass of the quark. $T$ is temperature.
$\alpha_{s}(k)$ is QCD running coupling constant defined as:
\begin{equation}
 \alpha_{s}(k)=\frac{4 \pi}{(33-2n_{f})\ln(1+k^{2}/\Lambda^{2})},
\end{equation}
in which $~\Lambda $ is QCD parameter. The value is taken to be~$ 0.15~$GeV. 
 $n_{f}$ is
degree of freedom of quark and gluon. 
So the interacting mean-field potential $ V_{conf}(k) $ is
now obtained with inclusion of two
loop correction
factor from simple confining potential obtained through
the Hamiltonian and it is modified from
the earlier potential. The modified potential is now
expressed through
the expansion of strong coupling constants of two loop factor within
the perturbation theory as~\cite{brambilla,melnikov,va}:
\begin{eqnarray}\label{3.18}
V_{\mbox{conf}}(k) &=& \frac{8 \pi}{k}\gamma_{q,g} ~ \alpha_{s}(k) T^{2} [1+\frac{\alpha_{s}(k)a_{1}}{4\pi} \nonumber \\ 
&+&\frac{\alpha_{s}^2(k)a_{2}}{16 \pi^2} ] - \frac{m_{0}^{2}}{2 k},
\end{eqnarray}

where the loop co-efficients $a_{1}$ and $a_{2}$ play the roles
for involving in the creation
of interacting potential. 
In the expression again, quark and gluon parametrization 
factors are defined as
$ \gamma_{q}=1/14 $ and $\gamma_{g}=~ (48 - 60)~ \gamma_{q}$.
These factors play many functional roles with the
creation of droplets. First it determines the critical droplet formation. 
It then increases the dynamics of QGP flow and it also enhances in the process
to transform QGP droplet to hadron droplets.
 Now the density of states in phase space with loop corrections
in the interacting potential is modified and obtained through 
a generalized Thomas- Fermi
model as~\cite{linde,singh}:
\begin{equation}
 \rho_{q,g} (k) = v/\pi^{2}[-V_{conf}(k)]^{2} \frac{dV_{conf}}{dk}~,
\end{equation}
or,

\begin{equation}\label{3.13}
\rho_{q, g} (k) =\frac{v}{\pi^2}[\frac{\gamma_{q,g}^{3}T^2}{2}]^{3} G^{6}(k)A, 
\end{equation}
where
\begin{eqnarray}
A&=&\lbrace 1+\frac{\alpha_{s}(k)a_{1}}{\pi}+\frac{\alpha_{s}^2(k)a_{2}}{\pi^2}\rbrace^{2}\nonumber \\
&\times&[ \frac{(1+\alpha_{s}(k)a_{1}/\pi+\alpha_{s}(k)^2a_{2}/\pi^2)}{k^{4}} \nonumber \\
&+&\frac{2 (1+2\alpha_{s}(k)a_{1}/\pi+3 \alpha_{s}(k)^2a_{2}/\pi^2)}{k^{2}(k^2+\Lambda^2)\ln(1+\frac{k^2}{\Lambda^2})}] 
\end{eqnarray}
and~ $v$ is the volume occupied by the QGP and
$~G^{2}(k)=4 \pi \alpha_{s}(k)(1+\frac{\alpha_{s}(k)a_{1}}{4 \pi})$.

\begin{figure*}[htb]
\centering
\includegraphics[width=7cm,clip]{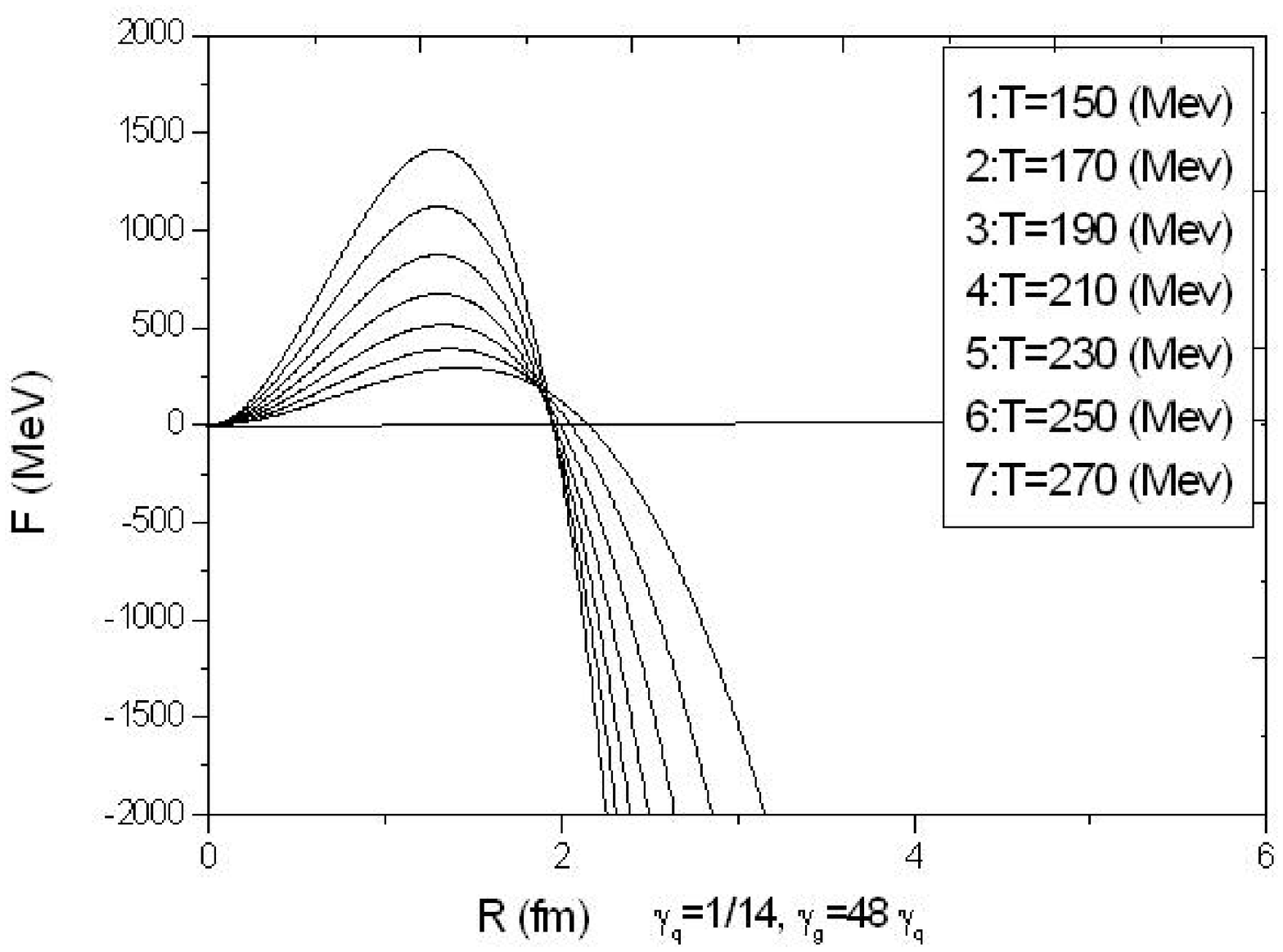}
\caption{The free energy vs.~R~at~$\gamma_{q}=1/14~$, $\gamma_{g}=48\gamma_{q}$ for various values of temperature.}
\label{fig-2}       
\end{figure*}

\begin{figure}[htb]
\centering
\includegraphics[width=7cm,clip]{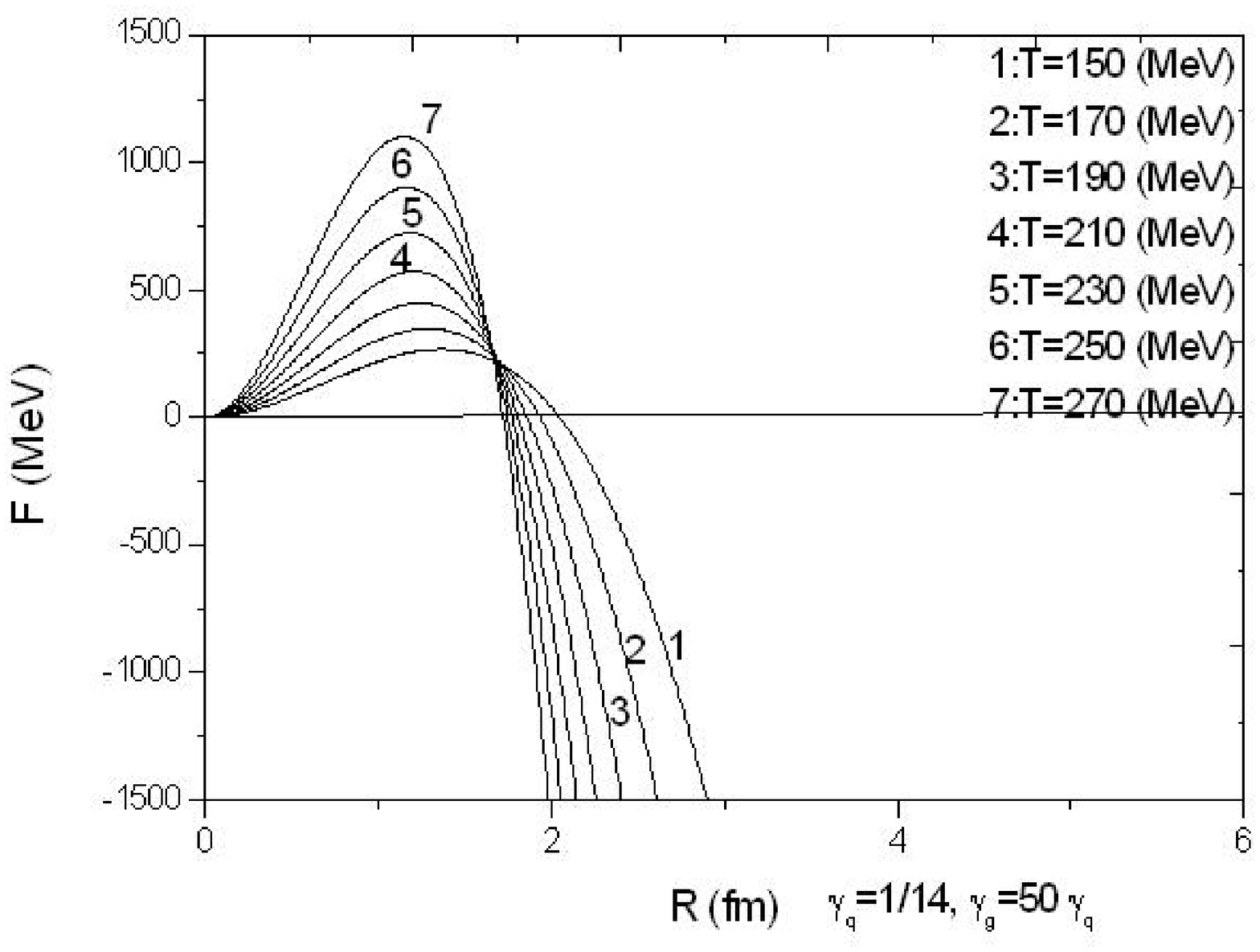}
\caption{ The free energy vs.~R~
at~$\gamma_{q}=1/14~$, $\gamma_{g}=50\gamma_{q}$ for the various values of temperature.}
\label{fig-3}       
\end{figure}

\begin{figure}[htb]
\centering
\includegraphics[width=7cm,clip]{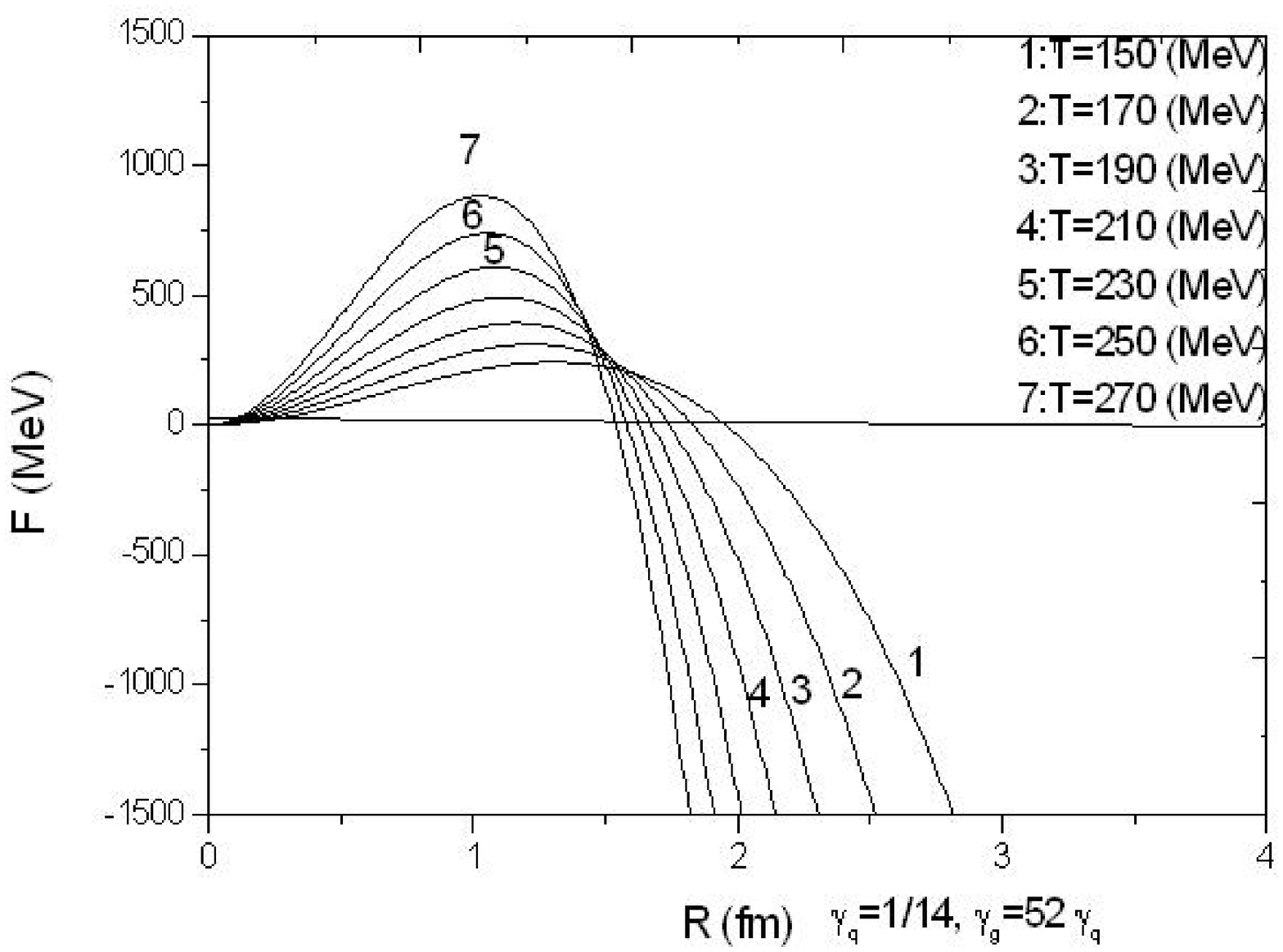}
\caption{The free energy vs.~R~
at~$\gamma_{q}=1/14~$, $\gamma_{g}=52\gamma_{q}$ for the various values of temperature.}
\label{fig-4}       
\end{figure}

\begin{figure}[htb]
\centering
\includegraphics[width=7cm,clip]{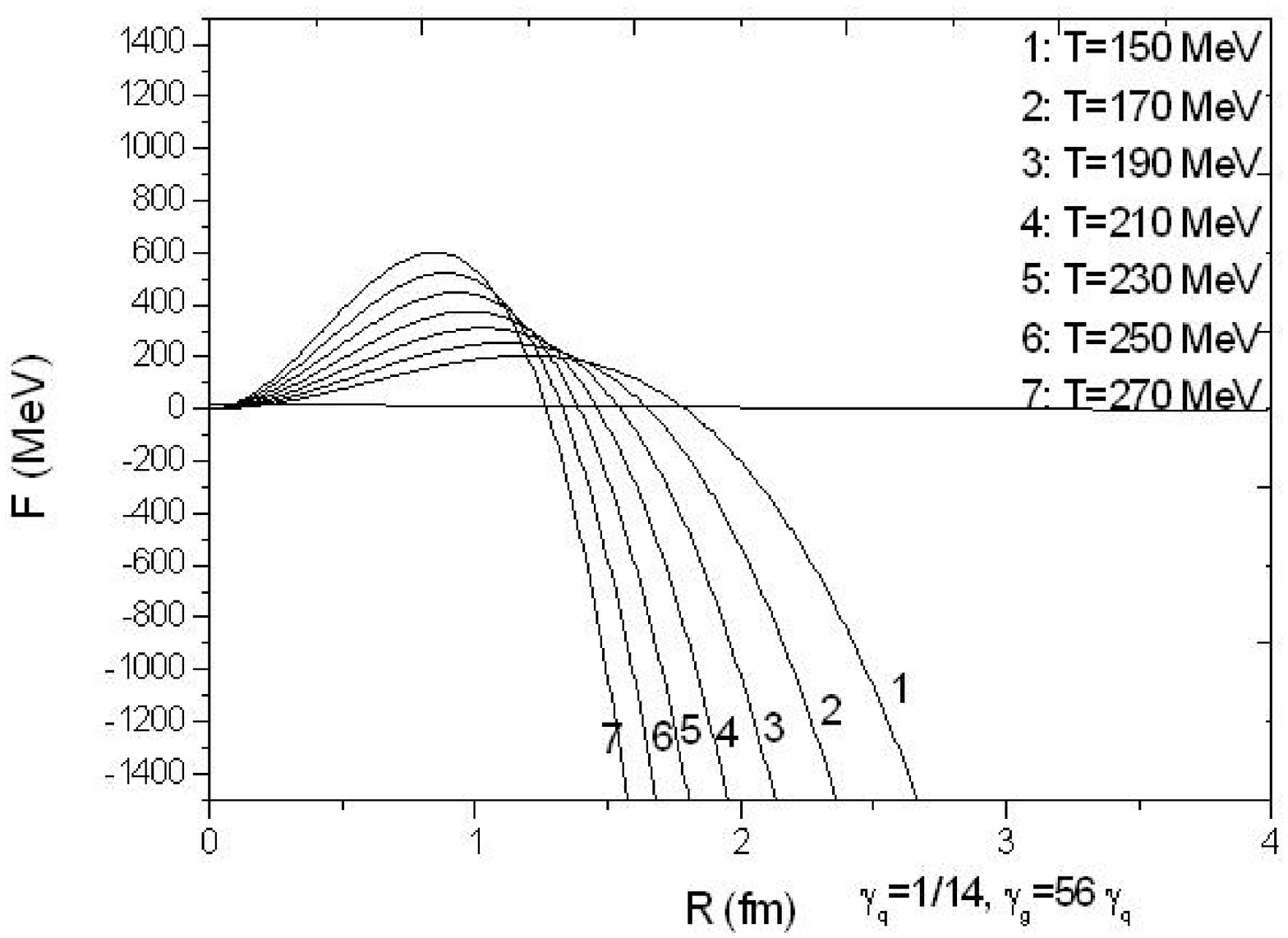}
\caption{The free energy vs.~R~
at~$\gamma_{q}=1/14~$, $\gamma_{g}=56\gamma_{q}$ for the various values of temperature.}
\label{fig-5}       
\end{figure}

\begin{figure}[htb]
\centering
\includegraphics[width=7cm,clip]{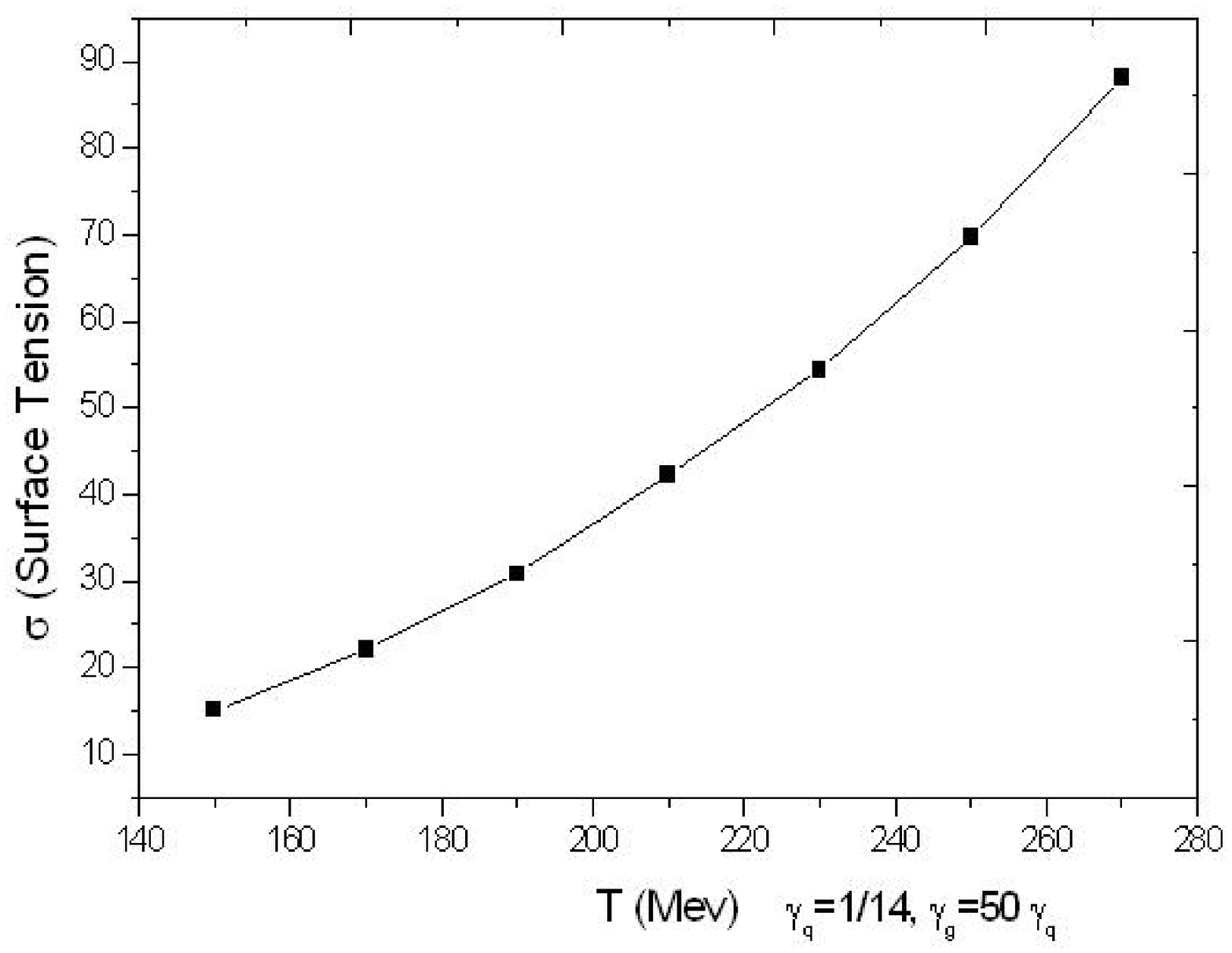}
\caption{The Surface Tension vs.~T~
at~$\gamma_{q}=1/14~$, $\gamma_{g}=50\gamma_{q}$.}
\label{fig-6}       
\end{figure}

\begin{figure}[htb]
\centering
\includegraphics[width=7cm,clip]{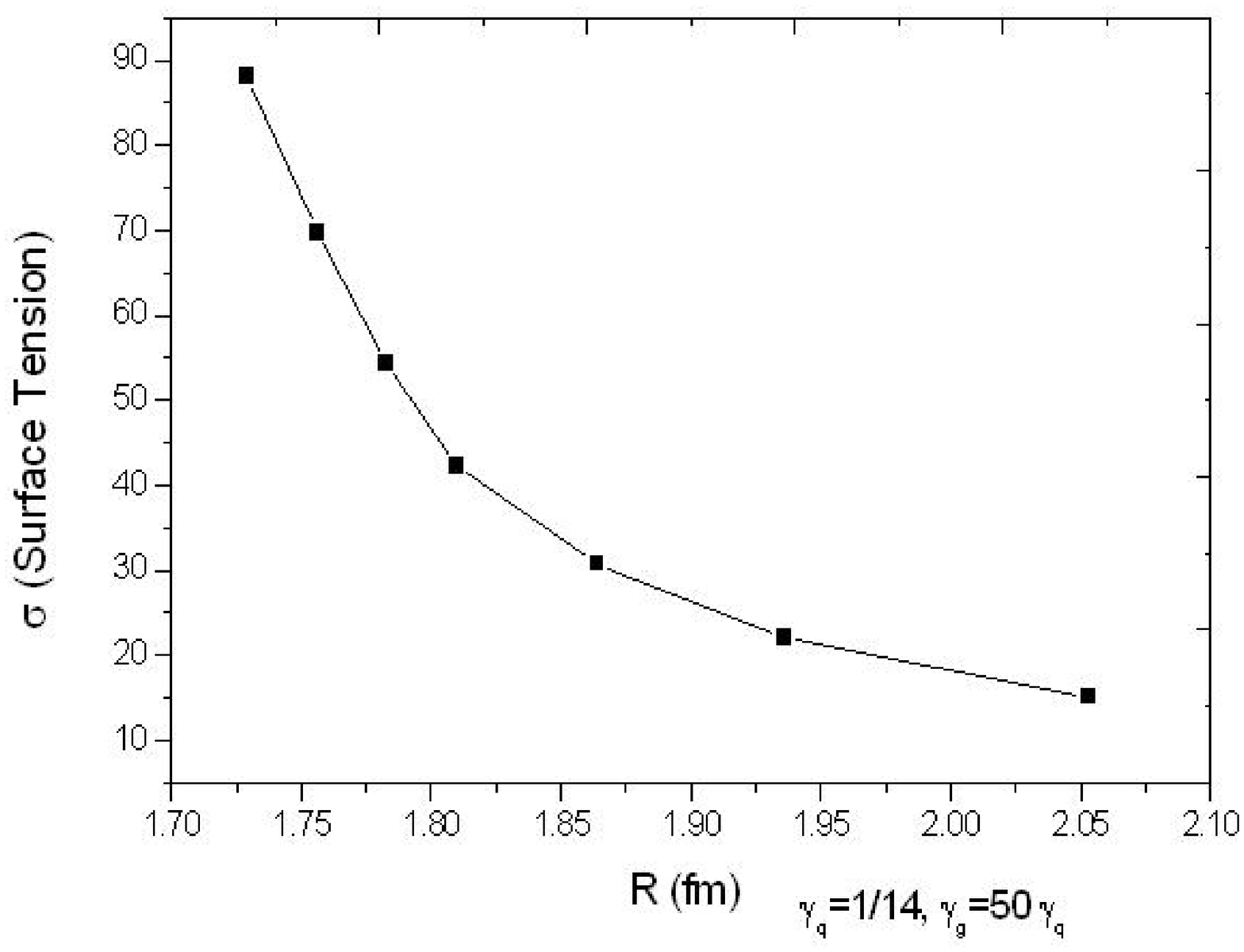}
\caption{The Surface Tension vs.~$R_{c}$~
at~$\gamma_{q}=1/14~$, $\gamma_{g}=50\gamma_{q}$.}
\label{fig-7}       
\end{figure}

\begin{figure}[htb]
\centering
\includegraphics[width=7cm,clip]{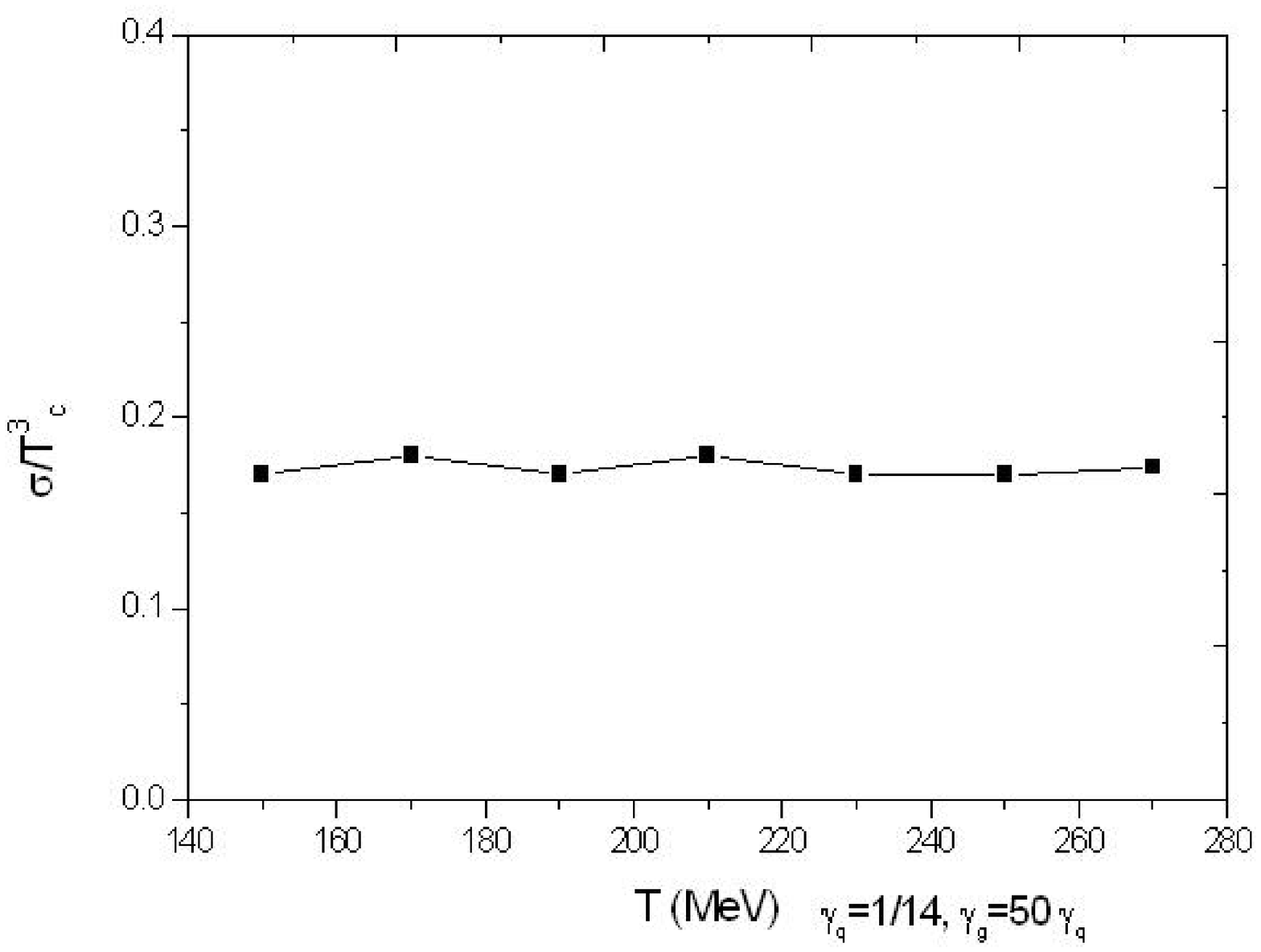}
\caption{The Surface Tension/$T_{c}^{3}$ vs.~T~
at~$\gamma_{q}=1/14~$, $\gamma_{g}=50\gamma_{q}$.}
\label{fig-8}       
\end{figure}

\section{The free energy evolution}
 The free energy
of quarks and gluons is defined in the following with the modified
density of states
as~\cite{mb}:
\begin{equation}\label{3.20}
F_i = -\eta T g_i \int dk \rho_{q,g} (k) \ln (1 +\eta e^{-(\sqrt{m_{i}^2 + k^2}) /T})~,
\end{equation}
where $\eta=+ve$ gives the bosonic particle and $\eta=-ve$ gives 
the contribution
from the fermionic particles.
The minimum potential cut off in terms of momentum is obtained as:
\begin{equation}\label{3.19}
V(k_{min})=(8 a_{1} \gamma_{g,q}N^{\frac{1}{3}} T^{2} \Lambda^4 /27 \pi^{2} )^{1/6},
\end{equation}
where $N=(4/3 )[12 \pi / (33-2 n_{f})]$.
The minimum cut off in
the model leads the integral to a finite value by avoiding
the infra-red divergence while taking
the magnitude of $\Lambda$ and
$T$ as of the same order of lattice QCD.
 $g_{i}$ is degeneracy factor (color and particle-antiparticle degeneracy)
which is $6$ for
quarks and~ $8$~ for gluons.
 The inter-facial energy
obtained through a scalar
Weyl-surface in Ramanathan et al. \cite{ramanathan,weyl} with a suitable
modification to take care of
the hydrodynamic effects is given as:
\begin{equation}\label{3.22}
  F_{interface}= \frac{1}{4}\gamma R^{2}T^{3}. 
\end{equation}
This interfacial energy is used to replace the bag energy of MIT model and 
it minimizes
the drawback produced by MIT model. $\gamma$ is root mean square value in
terms of
quark $\gamma_{q}$ and gluon parameter $\gamma_{g}$. 
The hadron free energy is~\cite{balian}

\begin{equation}\label{3.25}
F_{h} = (d_{i}T/2\pi^2 ) v \int_0^{\infty} k^2 dk \ln (1 - e^{-\sqrt{m_{h}^2 + k^2} / T}).
\end{equation}
where $d_{i}$ is the degeneracy factor for the different
light hadronic particles and $m_{h}$ is the light hadron corresponding masses.
Becuase we considered only light hadrons as they
are produced as maximum amount in the reaction plane. 
To calculate the total free energies, the
particle masses are taken as:
quark masses $m_u = m_d = 0 ~ MeV$
and $m_s = 0.15 ~ GeV$~ [11]. Now
we can compute the total modified free energy $F_{total}$ as,
\begin{equation}
 F_{total}=\sum_{i} F_{i}~+~F_{interface}~+~F_{h},
\end{equation}
~ where $i$ stands for $u$,~$d$
and $s$~quark and gluon. 
\section{Surface tension with effect of two loop correction}
The surface tension with two loop correction in the potential 
is calculated using the 
difference relation between free energy of QGP phase and the light element 
hadrons phase. It will indicate a sharp boundary between
hadron and quark phase, balancing the pressure and keeping
chemical equilibrium in the mixed phase which
is normally treated by the Gibbs condition of a simple bulk calculation 
of free energy.
However the calculation of surface tension
can be highly affected by the mixed phase in which large finite
size effects are included~\cite{yasu,dhar}. To exclude the effects of
finite size, we exclude the mixed phase system in the present  
calculation. So
the difference in energy of the two phases define critical phase 
transition of liquid drop model after neglecting the finite
size effects and shape contribution. It is therefore given as:
\begin{equation}
\Delta F =- \frac{4 \pi}{3}R^3[P_{had}(T)-P_{q,g}(T)] + 4 \pi R^2 \sigma
\end{equation}
where the first term represents pressure difference and second term represents
the contribution from the surface tension. The surface tension is calculated
by minimizing the above expression with respect to droplet size.
So, the surface tension formula is obtained as:
\begin{equation}
Rc=\frac{2 \sigma}{\Delta p}~ or~ \sigma =\frac{3 \Delta F}{4 \pi R_{c}^2}
\end{equation} where, $\Delta F$ is the change in the free energy and $R_{c}$
is the corresponding critical radius obtained at the transition point
from quark droplet to hadron droplet.

\section{Results:} The effects
 of QGP droplet formation with the inclusion of two loop
correction factor in the interacting mean-field potential is 
numerically calculated. Due to the inclusion of the two loop correction, 
the QGP droplet changes a lot from one loop
correction and without loop correction. The modifications in the droplet 
sizes are replicated in the figures showing in the free energies. 
The results in the free energies are modified by
the quark and gluon flow parameters involved in the lwo loop correction
as interacting parameter.
In Fig.$1$, we can see stability evolution of droplet at the particular
quark and gluon flow parametrization factor 
$~\gamma_{q}=1/14,~\gamma_{g}=48 \gamma_{q}$. There is a stable
droplet formation at all the temperatures forming the droplet size 
of $2.0~$fm radius with the effects of two loop correction. Now we keep on
increasing the gluon flow parameter fixing the quark flow parameter.
We get again the stable droplet 
for all the temperatures at another gluon flow parameter.  
The size of the droplet is found to be
around $R \le 2.0~$fm and gluon parametrization what we found is to be 
$\gamma_{g}=50\gamma_{q}$. It means that stability of droplet
is really observed for all
the different temperatures at the certain range of gluon parameters. 
In these two droplets we
obtain the free energy amplitude less than 
$2.0~GeV$ and at lesser gluon parameter the energy amplitude
is larger with large stable droplet size. In Fig.$3$ we further
increase the gluon
flow parameter.                              
We obtain a slightly stable droplet with the 
increase of gluon
flow parameter $\gamma_{g} < 52 \gamma_{q}$ and the amplitude of the free
energy is found to be lesser.
This indicates that adopting the flow parameter in the 
range $\gamma_{q}=1/14$
and ~$\gamma_{g}<=52 \gamma_{q}$, we obtained stability
of the QGP droplet and the amplitude of the free energies drop down with
increasing gluon flow parameter.
In Fig.$4$, we further increase gluon parameter up to
$\gamma_{g} = 56 \gamma_{q}$ then we observe unstable droplet 
formation as we increase the gluon
parameter from $\gamma_{g}=52 \gamma_{q}$ to higher 
value $\gamma_{g}=56 \gamma_{q}$. 
It indicates that instability starts forming 
from the gluon parameter $\gamma_{g}=52 \gamma_{q}$  with QGP 
droplet formation with the effect
of two loop correction in the potential. Such type of effects are obtained
earlier in one loop correction and the droplet sizes are bigger. With
the addition of two loop correction in the potential, the size of the droplet
decrease and stable droplet formation are tightly bound in comparison
to the earlier droplet formation of one loop correction.
It implies that no further stability is observed with the 
increasing gluon flow parameter.
So stable droplet formation is specially found in the range of
parametrization factor
 $48 \gamma_{q} \le \gamma_{g} \le 52 \gamma_{q}$ and the amplitude
of the free energy with the stability is modified by these quark and gluon
flow parameter.
\par We again calculate the surface tension at these particular quark and gluon
flow parameter where more stable 
droplets are obtained. Calculating the surface tension of droplet
is the characteristic feature of fluid 
to determine the stability of the droplet. On the basis of this character,
the stable droplet features are shown in Figs.$5-7$. The Fig.$5$
indicates the increasing strength of surface tension with increasing 
temperature  at the parametrization values of  $48 \gamma_{q} \le \gamma_{g} \le 52 \gamma_{q}$. In Fig.$6$, we again observe 
the decreasing order of surface tension with the increasing critical 
radius of droplets. As the size of droplet is smaller
we get larger surface tension so that QGP droplets are tightly bounded and 
more stable.  
Increasing  the size of the droplet the surface tension is bound to be lesser.
In Fig.$7$, we again plot the ratio of surface tension to the cube of
critical temperature showing constancy of $\sigma/T_c^3$ with the 
temperature~\cite{som}. It is to show the comparative result with the lattice
data. The result
is found to be $\sigma =0.173~T_c^3 $ which is almost near the lattice result
 $\sigma=0.2~T_c^3$~\cite{kaja,iwasaki,michael}
So the inclusion of two loop correction in the mean field potential
with these parametrizations improve and enhance the stability of QGP droplet. 

\section{conclusion:}
The results show the effects on the stability of droplet in the presence of
two loop correction in the mean field potential. 
The effects of the stability is increased when the droplet size decreases 
as indicated
by Fig(6). The size of droplet
is more affected by the gluon flow parameter. If the parameter is increased
beyond $\gamma_{g} \ge 52\gamma_{q}$ then unstable droplet starts forming 
and size of droplet is difficult to predict. In the range of the gluon
flow paramenters say $48 \gamma_{q}\le \gamma_{g} \ge 52\gamma_{q}$ the stable
droplets are formed and the stability is more in the case of two loop 
correction in comparison to the one loop correction~[9]. It indicates
that two loop correction with the dynamical flow parameter
can enhance the stable droplet formation. This is another possible 
indication that evolution of QGP fireball is steady dynamics depending on 
some kind of dynamical parameter which plays in forming the stable droplets.

\subsection{\bf Acknowledgments:}

The author thanks R. Ramanathan for useful discussions
and critical reading of the manuscript and thanks
the University
for providing research and development grants for this work successful.


\begin{thebibliography}{}
\bibitem{ali}
Khan A A {\it et al.} CP-PACS Collab: 2001 {\it Phys. Rev D} {\bf 63},
034502;
Karsch E, Peikert A and Laermann E, 2001 {\it Nucl. Phys. B} {\bf 605}, 579.
\bibitem{satz}
Satz H 1978 {\it CERN-TH-2590} 18pp; Karsch F, Laermann E, Peikert A, Schmidt Chand Stickan S 2001{\it Nucl. Phys. B} {\bf 94}, 411; Karsch F and Satz H, 2002{\it Nucl. Phys. A} {\bf 702}, 373.
\bibitem{mardor}
Mardor I, Svetitsky B, 1991{\it Phys. Rev. D} {\bf 44}, 878.
\bibitem{neergaard}
Neergaard G, Madsen J, 2000 {\it Phys. Rev. D} {\bf 62}, 034005.  
\bibitem{peshier}
Peshier A, K$\ddot{a}$mpfer B, Pavlenko O P and Soff G, 1994
{\it Phys. Lett. B} {\bf 337}, 235; Goloviznin V and Satz H, 1993 {\it Z. Phys. C} {\bf 57}, 671.
\bibitem{rama}
Ramanathan R, Mathur Y K, Gupta K K and Jha A K, 2004{ \it Phys. Rev. C} {\bf 70}, 027903.
\bibitem{fischler}
Fischler W, 1977 {\it Nucl. Phys. B} {\bf 129}, 157; Billoire A, 1980 {\it Phys. Lett B} {\bf 92}, 343.
\bibitem{smirnov}
Smirnov A V, Smirnov V A, Steinhauser M,
2008 {\it Phys. Lett. B} {\bf 668}, 293;
Smirnov A V, Smirnov V A and Steinhauser M,
2010 {\it Phys. Rev. Lett.} {\bf 104}, 112002.
\bibitem{singh}
Singh S S, Ramanathan R, 2014 {\it Prog. Th. Expt. Phys.} {\it 103D02}.
\bibitem{ramanathan}
Ramanathan R, Gupta K K, Jha A K and Singh S S, 2007 {\it Pram. J. Phys.} {\bf 68}, 757.
\bibitem{brambilla}
Brambilla N, Pineda A, Soto J and Vairo A, 2001 {\it Phys. Rev D} {\bf 63}, 014023.
\bibitem{melnikov}
Melnikov K and Yelkhovsky A, 1998 {\it Nucl. Phys. B} {\bf 528}, 59;
Hoang A H, 1999 {\it Phys. Rev D} {\bf 59}, 014039.
\bibitem {va}
V. A. Yerokhin, P. Indelicato, V. M. Shabaev, 2007 {\it Can. J. Phy} {\bf 85}, 251. 
\bibitem{mb}
Christiansen M B and Madsen J, 1997{\it J. Phys. G: Nucl. Par. Phys} {\bf 23}, 2039;
Elze H T and Greiner W, 1986{\it Phys. Lett. B} {\bf 179}, 385.
\bibitem{linde}
Linde A D, 1983{\it Nucl. Phys. B} {\bf 216}, 421; Fermi E, 1928{\it Zeit F. Physik} {\bf 48}, 73; Thomas L H, 1927{\it Proc. Camb. Phil. Soc.} {\bf 23}, 542;Bethe H A, 1937 {\it Rev. Mod. Phys.} {\bf 9}, 69.
\bibitem{weyl}
Weyl H, 1911 {\it Nachr. Akad. Wiss Gottingen} 110.
\bibitem{balian}
Balian R and Block C, 1970 {\it Ann. Phys. (NY)} {\bf 60}, 401.
\bibitem{yasu}
Yasutake N, Maruyama T, Tatsumi T, 2011{\it J. Phys. Conf. Series} {\bf 312} 042027.
\bibitem{dhar}
D. S. Gosain and S. S. Singh, 2014 {\it Intn. J. Theo. Phys} {\bf 53} 2688.
\bibitem{som}
Singh S S, Gupta K K, Jha A K, 2014 {\it Int. J. Mod. Phys. A} {\bf 29} 1450097.
\bibitem{kaja}
Kajantie K, K\"{a}rkk\"{a}inen L and Rummukainen K, 1990 {\it Nucl. Phys. B} {\bf 333} 100.
\bibitem{iwasaki}
Iwasaki Y, Kanaya K, K\"{a}rkk\"{a}inen L, Rummukainen K and Yoshi\'{e} T, 1994 {\it Phys. Rev. D} {\bf 49} 3540. 
\bibitem{michael}
Christiansen M B and Madsen J, 1996 {\it Phys. Rev. D} {\bf 53} 5446. 
\end{thebibliography}
\end{document}